\documentstyle[preprint,epsfig,epsf,eqsecnum,aps]{revtex}

\begin{document}

\thispagestyle{empty}
\setcounter{page}{1}

\title{LIFETIME EFFECTS IN COLOR SUPERCONDUCTIVITY AT WEAK COUPLING}

\author{Cristina Manuel\footnote{E-mail: Cristina.Manuel@cern.ch} }

\address{Theory Division, CERN, CH-1211 Geneva 23, Switzerland}

\maketitle

\thispagestyle{empty}
\setcounter{page}{0}

\begin{abstract}
$\!\!$Present computations of the gap of color superconductivity
in weak coupling  assume that the quarks which participate
in the condensation process are infinitely long-lived.
However, the quasiparticles in a plasma are characterized by having
a finite lifetime. In this article we take into account this fact
to evaluate its effect in the computation of the
color gap. By first considering
the Schwinger-Dyson equations in weak coupling, when one-loop
self-energy corrections are included, a general gap
equation is written in terms of the spectral densities of the
quasiparticles. To evaluate  lifetime effects, we then model
the spectral density by a Lorentzian function. We argue that 
the decay of the quasiparticles limits their efficiency to
condense. The value of the gap at the Fermi surface is then
reduced. To leading order, these lifetime effects 
 can be taken into account by replacing the coupling constant
of the gap equation by a reduced effective one.
\end{abstract}

\vfill

\hfill\break
\hbox to \hsize{CERN-TH/2000-160} 
\hbox to \hsize{June/2000}
\vskip-12pt
\eject

\baselineskip= 20pt
\pagestyle{plain}
\baselineskip= 20pt
\pagestyle{plain}

\section{Introduction}

In QCD at high baryonic density, when asymptotic freedom implies
that the   interactions are weak, matter behaves as a color
superconductor \cite{Bailin:1984,Alford:1998,Rapp:1998}. 
This is a consequence of Cooper's theorem,
as any attractive interaction occurring close to the Fermi
surface makes the system unstable to the formation of particle
pairing. In QCD the attractive interaction is provided by
one-gluon exchange between quarks in a color  antisymmetric ${\bar 3}$
channel.

In the weak coupling regime it is possible to compute
the value of the quark-quark condensate
\cite{Son:1999,Schafer:1999b,Pisarski:2000bf,Pisarski:2000tv,Hong:2000tn,Hong:2000fh}.
The condensation process
is dominated by the exchange of very soft magnetic gluons \cite{Son:1999}, 
which are
dynamically screened by medium effects. A careful analysis
of the gap equation, as arising from the Schwinger-Dyson equations,
was done in Refs. \cite{Schafer:1999b,Pisarski:2000bf,Pisarski:2000tv},
and the value of the gap  was determined up to a
constant of order one.  The inclusion of
the Meissner effects in the gluon exchange processes
introduces a slight correction to the value of the gap \cite{Evans:2000at}.
The gap can also be computed  directly from the scattering
matrix, looking into the instability in the proper vertex function
\cite{Brown:2000aq,Brown:2000yd}. From this approach, the undetermined
constant of the gap was  computed.

In this article we study   the gap equation  when
the one-loop fermion self-energy corrections are taken into
account in the Schwinger-Dyson equations.
 The propagation properties of the quasiparticles in
the dense medium are not the same as in vacuum. A
more accurate computation of the gap necessarily needs to
take into account this fact. One of the 
most relevant characteristics of the quasiparticles 
in the medium is that these have a finite lifetime
\cite{LeBellac:1997kr,Vanderheyden:1997bw,Manuel:2000mk}.
These quasiparticles scatter with the quarks inside the Fermi sea,
and they decay. In all the previous estimates of the color gap
these effects have been ignored to leading order.
What is the effect of a finite lifetime in the gap equation?
Physically, one would expect that the decay of the quasiparticles
will limit their efficiency to participate in the condensation
process. For those quasiparticles whose lifetime is very short
(and this is the case for those who are far away from the Fermi
surface), their chance to participate in  Cooper pairing 
 is rather small. Taking  properly into account the
damping rate of the quasiparticles should then allow us to
distinguish which are the modes which participate in the
condensation. This should then provide a
physical ultraviolet cutoff in the
gap equation. This is actually the most relevant
effect of including the damping rate in the gap equation for 
BCS superconductivity in weak coupling \cite{Morel}.
 Here we want to estimate the lifetime effects in the framework
of   color superconductivity.

This paper is structured as follows. In Section II we write the
Schwinger-Dyson equations of the superconducting phase of QCD in
the general case when one-loop fermion self-energy corrections
are not neglected. We will however neglect vertex corrections
throughout our analysis. Those equations are the generalization
to color superconductivity of the equations that Eliashberg considered
for BCS superconductivity \cite{Eliashberg}.
 In the most general case, these equations
are very hard to solve. However, in weak coupling  the
one-loop fermion self-energies are the same, up to corrections of the
order of the squared of the condensate, in the superconducting and
normal phases of the system. Therefore, one can include the 
one-loop self-energy corrections of the normal phase of the system
in the Schwinger-Dyson equations.
 With this last approximation, the
gap equation can be solved in an easier way.
In Section III, and for purposes of comparison, we briefly review 
the computation  of the gap  in the free quasiparticle
approximation  of Ref. \cite{Pisarski:2000tv}.
 We will only consider the case
of  $N_f=2$ quark massless flavors throughout this article.
A generalization of our results  to a different number of quark flavors
is rather straightforward. In Section IV, we
derive the gap equation when the one-loop self-energy corrections
are included. We give a final expression of the gap equation
written in terms of the one-loop spectral density of the quasiparticles
at an arbitrary small value of the temperature $T$.
To evaluate the lifetime effects at $T=0$,
we model this
spectral density by a Lorentzian function. We then make a rough
estimate of  the correction to the value of the gap
due to the inclusion of these lifetime effects. Since 
the damping rate has a linear dependence on the quasiparticle
energy \cite{LeBellac:1997kr,Vanderheyden:1997bw,Manuel:2000mk},
 rather than quadratic as in BCS superconductivity,
we then find that, to leading order, the effect of the damping rate
in the gap is to  reduce the effective coupling constant
that the quasiparticles close to the Fermi surface see through the
condensation process. We present our conclusions in Section
V.

\section{Schwinger-Dyson Equations in the Superconducting Phase
of QCD at  weak coupling}

\subsection{Eliashberg equations}

Here we will follow 
the Nambu-Gorkov (NG) formalism to study the superconducting
phase of QCD at very high baryonic density.
In the  NG  basis
\begin{equation}
\label{eq.A1}
\Psi = \left( \begin{array}{c}
               \psi \\ \psi_C
             \end{array} \right) \ , \qquad
{\bar \Psi} = \left({\bar \psi}, {\bar \psi}_C \right) \ ,
\end{equation}
where $\psi_C (x) = C {\bar \psi}^{T}(x)$ is the charge-conjugate spinor,
the Schwinger-Dyson (SD) equations read
\begin{equation}
\label{eq.A2}
S_{\alpha \beta} = S^{(0)}_{\alpha \beta} +  S^{(0)}_{\alpha \gamma} \Sigma_{\gamma \delta} 
S_{\delta \beta} \ ,
\end{equation}
where the Greek indices  refer to the upper/down components of 
the NG fields. Every term in the NG matrix is also a matrix in color,
flavor and Dirac space. We will suppress throughout
 the color and flavor indices,
unless necessary to avoid confusion. The free propagators 
$S^{(0)}_{\alpha \beta}$ are
\begin{equation}
\label{eq.A3}
S_{11}^{(0)}(P) =\frac{1}{{P\llap{/\kern1pt}} + \mu \gamma_0 - m} 
\ , \qquad 
S_{22}^{(0)}(P) = \frac{1}{{P\llap{/\kern1pt}} - \mu \gamma_0 -m} \ ,
\end{equation}
while $S_{12}^{(0)} = S_{21}^{(0)} =0$.
 $\Sigma_{\gamma \delta}$ denote the self-energy corrections. Notice
that
$\Sigma_{21}$ and $\Sigma_{12}$ imply the creation and annihilation,
respectively, 
of a condensate, while the terms $\Sigma_{11}$, $\Sigma_{22}$ do not.
Vertex corrections to the SD equations
 will be omitted at the order of approximation
we are working.

The values of the one-loop propagators can be obtained from the 
SD equations considered above.
After solving the set of coupled equations for $S_{21}$ and $S_{11}$,
one finds
\begin{equation}
\label{eq.A4}
S_{21} = - \left( 1 - S_{22}^{(0)} \Sigma_{22} \right)^{-1} \, S_{22}^{(0)} \Sigma_{21}
S_{11} \ ,
\end{equation}
and
\begin{equation}
\label{eq.A5}
S_{11} = \left(1 - S_{11}^{(0)} \Sigma_{11} - S_{11}^{(0)} \Sigma_{12}
\left( 1 + S_{22}^{(0)} \Sigma_{22} \right)^{-1} \, S_{22}^{(0)} \Sigma_{21}
 \right)^{-1} S_{11}^{(0)} \ .
\end{equation}

The gap equation is now derived by demanding  the value of the condensate
be the same as its one-loop correction. Using the imaginary time formalism,
where $K=(k_0,{\bf k})$, and $k_0 = -i \omega_n$, where $\omega_n$
is a fermionic Matsubara frequency\footnote{Our notations and conventions  are 
almost the same as in Ref. \cite{Pisarski:2000tv}, the only change being the notation for
the propagators. To compare, just notice that
 $S^{(0)}_{11/22} = G_{0}^{\pm}$
and $\Sigma_{21} = \Phi^+$, $\Sigma_{12} = \Phi^- = 
\gamma_0 (\Phi^+)^\dagger \gamma_0$.} 
\begin{equation}
\label{eq.A6}
\Sigma_{21} (K) = g^2 \frac{T}{V} \sum_{Q} {\bar \Gamma}_a^\mu \Delta_{\mu \nu}^{ab} (K-Q)
S_{21}(Q) \Gamma^\nu _b \ ,
\end{equation}
where $g$ is the coupling constant,
$\Delta^{ab}_{\mu \nu}$ is the gluon propagator, and the vertices
are $\Gamma^\mu _a \equiv T_a \gamma^\mu$ and ${\bar \Gamma}_a^\mu
\equiv C (\Gamma^\mu _a)^T C^{-1} \equiv - \gamma^\mu T^T_a$, and
$T_a = \lambda_a/2$, where $\lambda_a$ are the Gell-Mann matrices.

In a self-consistent treatment of the system, one obtains the
fermion self-energy as 
\begin{equation}
\label{eq.A7}
\Sigma_{11} (K) = g^2 \frac{T}{V} \sum_{Q} {\Gamma}_a^\mu \Delta_{\mu \nu}^{ab} (K-Q)
S_{11}(Q) \Gamma^\nu _b \ . 
\end{equation}

Notice that if one ignores the functions $\Sigma_{11}$ 
and $\Sigma_{22}$,  Eq. (\ref{eq.A6})  reduces to the gap
equation already considered in the literature.

The gluon propagator of Eqs. (\ref{eq.A6}-\ref{eq.A7}) has also to be
computed self-consistently. The one-loop gluon self-energy corrections
are given by (see \cite{Rischke:2000qz,Rischke:2000ra} for an
the explicit computation at leading order)
\begin{eqnarray}
\label{eq.A7bis}
\Pi^{\mu \nu}_{ab} (P) & = & \frac{g^2}{2} \frac{T}{V} \sum_{K}
{\rm Tr}_{s,c,f} \left[ {\Gamma}_a^\mu S_{11} (K) {\Gamma}_b^\nu 
S_{11}(K-P) + {\bar \Gamma}_a^\mu S_{22} (K) {\bar \Gamma}_b^\nu 
S_{22}(K-P) \right.  \\
& + &  \left. {\Gamma}_a^\mu S_{12} (K) {\bar \Gamma}_b^\nu 
S_{21}(K-P) + {\bar \Gamma}_a^\mu S_{21} (K) {\Gamma}_b^\nu
S_{12}(K-P) \right] \ . \nonumber
\end{eqnarray}

\subsection{Weak coupling limit and the Meissner effect}

Equations (\ref{eq.A6}-\ref{eq.A7bis}) are the generalization of 
the Eliashberg equations \cite{Eliashberg}
to color superconductivity. They form a set of coupled integral equations,
which are extremely hard to solve. In the weak coupling limit, however,
this set of equations can be decoupled and simplified, 
due to the large hierarchy of scales in the theory.

We will first make the approximation that
 $\Sigma_{11}$ is given, up to corrections
of the order of the squared of the condensate, by the value it
would take in the normal phase of the system. 
That is, one can approximate Eq. (\ref{eq.A7}) by replacing $S_{11}$
by $S_{11}^{(0)}$. 

A similar approximation can be made in the gluon propagator. One can
use the gluon propagator in Eqs. (\ref{eq.A6}-\ref{eq.A7}) 
with the value it would take in the normal phase of the system,
that is, in the hard dense loop approximation (HDL) 
\cite{Pisarski:1989vd,Manuel:1996td}, neglecting to leading order
the Meissner effect.

The reason why one can neglect the Meissner effect in the computation
of $\Sigma_{21}$ was first explain in Ref. \cite{Son:1999} and afterwards
confirmed in Refs. 
\cite{Schafer:1999b,Pisarski:2000bf,Pisarski:2000tv,Hong:2000tn,Hong:2000fh}.
The scattering processes of the quarks close to the Fermi surface,
the ones responsible of the Cooper's instability,
are dominated by small angle (or collinear) scattering. These processes
are mediated by the interchange of soft gluons. For soft gluons,
the value of the polarization tensor Eq. (\ref{eq.A7bis}) is dominated
by the HDL contribution (see Refs. \cite{Rischke:2000qz,Rischke:2000ra}),
while the Meissner effect is subleading.

The reason why one neglect the Meissner effect in the computation
of $\Sigma_{11}$ close to the Fermi surface is essentially the
same as for $\Sigma_{12}$. The imaginary part of $\Sigma_{11}$ evaluated
close to the Fermi surface  describes the scattering of a quark
close to the Fermi surface
with the quarks inside the Fermi sea.  This
process is dominated by collinear scattering, with the exchange of
 soft magnetic gluons which have space-like momenta
\cite{LeBellac:1997kr,Vanderheyden:1997bw,Manuel:2000mk}.
The Landau damping effect of magnetic gluons is then fully dominant
with respect to the Meissner effect.

In the approximation of using the HDL gluon propagator in 
Eqs. (\ref{eq.A6}-\ref{eq.A7}), these last integral 
equations are  decoupled  from Eq. (\ref{eq.A7bis}). 
This allows for  analytical computations of corrections
of order ${\cal O}(g^2)$ of the value of the gap,
which would be otherwise impossible. Only a numerical
analysis, as the one carried out in Ref. \cite{Evans:2000at},
can estimate the corrections introduced by neglecting
the Meissner effect. This is beyond the scope of this
paper.

\section{The gap equation in the free quasiparticle approximation}

In this section we review the computation  of the gap 
in the case where we consider that the quasiparticles which form
the Cooper pairs are not further affected by medium effects. We will follow
closely the computation of Ref. \cite{Pisarski:2000tv},
and in the following section
we will simply comment on how this computation is modified when the
self-energy corrections are included.

From now on, we will only treat the case of two quark flavors in the massless
limit. In this case both the color, flavor and Dirac structure of the gap equation
simplify drastically. 
By restoring color ($i,j,k=1,2,3$) and flavor ($f,g=1,2$)
 indices,  $\Sigma_{21}^{ij,fg}= \epsilon_{fg} \epsilon_{ijk} \Phi^+ _k$,
and one may take $\Phi^+_k = \Phi^+ \delta_{k3}$.
The Dirac structure of the condensate is 
\begin{equation}
\label{eq.A8}
 \Phi^+ (K) = \sum_{h=r,l} \sum_{e =\pm} \phi^{e}_h (K) {\cal P}_h \Lambda_{\bf k}^e \ ,
\end{equation}
where ${\cal P}_h$, and $\Lambda_{\bf k}^e$ are chirality and energy projectors,
respectively
\begin{equation}
\label{eq.A9}
{\cal P}_r = \frac{1 + \gamma_5}{2} \ , \qquad
 {\cal P}_l = \frac{1 - \gamma_5}{2} \ , \qquad
\Lambda_{\bf k} ^{\pm} = \frac{1    \pm  \gamma_ 0 {\bf \gamma} \cdot
{\bf \hat k}  }{2 } \ .
\end{equation}

The functions $\phi^+_h$ and $\phi^-_h$ are commonly known as
the gap and the antigap, respectively. We will focus on the equation for 
the gap. The antigap has not even been  computed to leading order.

In the presence of the above condensate the color group $SU(3)_c$ is
broken to $SU(2)_c$. The gap equation can  be simplified if
one takes into account that the gauge field modes which  contribute the most
to the integral are very ``soft'' \cite{Son:1999}. Then,  
one can take the gluon propagators in the
HDL approximation \cite{Pisarski:1989vd,Manuel:1996td},  neglecting  to leading order  
the Meissner effect. Also, one can drop the chirality index
$h$, as the equations for the right- and left- handed gaps 
at very high density are
identical and decoupled. In Coulomb gauge, the gap equation
reduces to \cite{Pisarski:2000tv}
\begin{eqnarray}
\label{eq.A10}
\phi^+(K) & = & \frac 23 g^2 \frac{T}{V} \sum_{Q} \left\{
\frac{\phi^+(Q)}{q_0^2 - \left(|{\bf q}| -  \mu \right)^2
-|\phi^+|^2 } \left[ \Delta_L (K-Q)
\frac{1 + {\bf \hat k} \cdot{\bf \hat q}}{2} \right. \right. \\
& +& \left. \left. \Delta_T (K-Q) \left(- \frac{3 - {\bf \hat k} \cdot{\bf \hat q}}{2} +
\frac{1 + {\bf \hat k} \cdot{\bf \hat q}}{2}
 \frac{(k-q)^2}{({\bf k} -{\bf q})^2} \right) \right] \right. \nonumber \\ 
&+& \frac{\phi^{-}(Q)}{q_0^2 - \left(|{\bf q}| +  \mu \right)^2 -|\phi^-|^2
} \left[ \Delta_L (K-Q)
\frac{1 + {\bf \hat k} \cdot{\bf \hat q} }{2} \right.  \nonumber \\
& +& \left. \left. \Delta_T (K-Q) \left(- \frac{3 - {\bf \hat k} \cdot{\bf \hat q}}{2} +
\frac{1 + {\bf \hat k} \cdot{\bf \hat q}}{2} \frac{(k-q)^2}
{({\bf k} -{\bf q})^2} \right) \right] \right\} \ ,  \nonumber 
\end{eqnarray}
where $\Delta_L, \Delta_T$ are the longitudinal and
transverse HDL propagators, respectively. In the above formula,
the gauge dependent pieces of the gluon propagator have been omitted,
since 
to leading order, they do not contribute in the determination of the
gap.  Also, the contribution of the quasi-antiparticles in the
above integral is very much suppressed, and it can be neglected.
If one evaluates the gap on-shell, and very close to the Fermi
energy, the integral is dominated by the contribution of
 very soft Landau-damped magnetic gluons. 
When the static electric gluons contribution is also taken into
account, one then arrives to  the integral equation
\begin{equation}
\label{eq.A11}
\phi_{k} = \frac{\bar g^2}{2} \int^{\delta}_0 \frac{ d (q-\mu)}{\epsilon_q}
\left[\ln{\left( \frac{\mu^2 b^2}{|\epsilon^2_q  -\epsilon^2_k|}\right)}
 \right] \phi_{q} \ ,
\end{equation}
where $\epsilon_q = \sqrt{\left(|{\bf q}| - \mu \right)^2 +
|\phi^+|^2}$,  $\phi_q \equiv \phi^+_q \equiv \phi^+(\epsilon_q,{\bf q})$,
and $\bar g= g/(3 \sqrt{2} \pi)$, and 
\begin{equation}
\label{eq.A12}
b =  256 \pi^4 (\frac{2}{N_f g^2})^{5/2} b'_0 \ ,
\end{equation}
where $b'_0$ is a constant of order one.
The integral in (\ref{eq.A11}) is limited to be around the Fermi
surface by introducing explicitly a cutoff $\delta \ll \mu$. The
final value of the gap finally does not depend on $\delta$.
The solution of the above equation for $k \sim \mu$ gives the value of
the gap $\phi_0$  to leading order in $g$
\begin{equation}
\label{eq.A13}
\phi_0 \sim 2 \, \frac{b_0}{g^5}\, \mu \exp{ \left(-\frac{\pi}{2 \bar g}\right)}
\left[1 + {\cal O}(\bar g) \right]  \ ,
\end{equation}
where $b_0 = g^5 b$.
In the following sections
we will see how lifetime effects of the quasiparticles
modify the above result.

\section{The gap equation including fermion self-energy corrections}

In this section we study how the self-energy corrections
to the quark propagators modify the gap equation, when one works in the
weak coupling limit. To that end, we first compute in the weak
coupling approximation the value of
$S_{21}$.

It is very convenient to project all the propagators and self-energy
corrections into the positive and negative energy contributions.
Thus
\begin{mathletters}
\label{eq.A14}
\begin{eqnarray}
\label{pr1}
[S_{11}^{(0)}]^{-1}(P) & = &  
  {P\llap{/\kern1pt}} + \mu \gamma_0 = 
\gamma_0 \Lambda_{\bf p}^+ \left(p_0 +\mu - |{\bf p}| \right) +
\gamma_0 \Lambda_{\bf p} ^- \left(p_0 + \mu + |{\bf p}| \right)  \ ,  
\\ 
\label{pr2}
[S_{22}^{(0)}]^{-1}(P) & = &  {P\llap{/\kern1pt}} - \mu \gamma_0 =
\gamma_0 \Lambda_{\bf p}^+ \left(p_0 - \mu - |{\bf p}| \right) +
\gamma_0 \Lambda_{\bf p} ^- \left(p_0 - \mu + |{\bf p}|  \right)
 \ , 
\end{eqnarray}
\end{mathletters}
$\!\!$and $\Sigma_{\alpha \beta}(P)  =  \sum_{e = \pm} 
\gamma_0 \Lambda_{\bf p}^{e} \Sigma^e_{\alpha \beta} (P)$, where
$\alpha=\beta=1,2$. Notice that in 
the weak coupling approximation we are considering,
both the right- and left-handed quarks get the same one-loop
corrections. 

We first compute $S_{11}$. We will approximate Eq. (\ref{eq.A5}) by
\begin{equation}
\label{eq.A16}
S_{11} \approx \left(1 - S_{11}^{(0)} \Sigma_{11} - S_{11}^{(0)} \Sigma_{12}
 \, S_{22}^{(0)} \Sigma_{21}
 \right)^{-1} S_{11}^{(0)} \ ,
\end{equation}
neglecting one of the quark self-energy corrections which multiplies
the condensate.  
With these approximations, one then reaches to
\begin{eqnarray}
\label{eq.A17}
S_{11} (P) &=& \sum_{h=r,l} {\cal P}_{ h} \left(
\Lambda_{\bf p}^+ \gamma_0 \frac{p_0 -\mu + |{\bf p}|}{
p_0^2 - ( |{\bf p}| - \mu)^2 -
|\phi_h^+|^2 - \left(p_0 -\mu + |{\bf p}| \right) \Sigma_{11}^+(P)
 }  \right. \\ 
 & +& \left. \Lambda_{\bf p}^- \gamma_0 \frac{p_0 -\mu - |{\bf p}|}{
p_0^2- (|{\bf p}| + \mu)^2
- |\phi_h^-|^2 - \left(p_0 -\mu - |{\bf p}| \right) \Sigma_{11}^-(P)
 } \right) \ ,\nonumber
\end{eqnarray}  
and 
\begin{eqnarray}
\label{eq.A18}
S_{21} = - \sum_{h=r,l} {\cal P}_{-h} \left(
\frac{ \Lambda_{\bf p}^- \phi^+_h (P)}{\left(
p_0 -\mu + |{\bf p}| - \Sigma_{22}^-(P)
 \right) \left(p_0 +\mu - |{\bf p}| - \Sigma_{11}^+(P) \right) -
|\phi_h^+|^2}
\right. \\
\left.
\frac{ \Lambda_{\bf p}^+ \phi^-_h (P)}{\left(
p_0 -\mu - |{\bf p}| - \Sigma_{22}^+(P)
 \right) \left(p_0 +\mu + |{\bf p}| - \Sigma_{11}^-(P) \right) -
|\phi_h^-|^2}
\right. \ .
\nonumber
\end{eqnarray}
Notice that in  Eq. (\ref{eq.A18}) , we have also neglected terms in
the denominators of order 
\begin{equation}
\label{eq.A19}
\frac{\Sigma_{22}^{\mp} |\phi^{\mp}_h|^2}{p_0 - \mu \pm |{\bf p}|} \ ,
\end{equation}
which give subleading corrections to the gap equation.

Then, with this fermion propagator, the gap equation  reads
(we drop the chirality index from now on)
\begin{eqnarray}
\label{eq.A20}
\phi^+(K) & = & \frac 23 g^2 \frac{T}{V} \sum_{Q} \left\{
\frac{\phi^+(Q)}{\left(
q_0 -\mu + |{\bf q}| - \Sigma_{22}^-(Q)
 \right) \left(q_0 +\mu - |{\bf q}| - \Sigma_{11}^+(Q) 
  \right) -|\phi^+|^2} \right.  \\
& \times&  \left[ \Delta_L (K-Q)
\frac{1 + {\bf \hat k} \cdot{\bf \hat q}}{2}  
 +  \Delta_T (K-Q) \left(- \frac{3 - {\bf \hat k} \cdot{\bf \hat q}}{2} +
\frac{1 + {\bf \hat k} \cdot{\bf \hat q}}{2}
 \frac{(k-q)^2}{({\bf k} -{\bf q})^2} \right) \right]  \nonumber \\ 
&+& \frac{\phi^{-}(Q)}{\left(
q_0 -\mu - |{\bf q}| - \Sigma_{22}^+(Q)
 \right) \left(q_0 +\mu + |{\bf q}| - \Sigma_{11}^-(Q)  \right) -|\phi^-|^2 
} \nonumber
 \\
& \times& \left.
\left[ \Delta_L (K-Q)
\frac{1 + {\bf \hat k} \cdot{\bf \hat q} }{2} 
 + \Delta_T (K-Q) \left(- \frac{3 - {\bf \hat k} \cdot{\bf \hat q}}{2} +
\frac{1 + {\bf \hat k} \cdot{\bf \hat q}}{2} \frac{(k-q)^2}
{({\bf k} -{\bf q})^2} \right) \right] \right\} \ . \nonumber
\end{eqnarray}

Taking into account that the relation between the self-energy corrections
to the fermion fields and charge-conjugate fields is given by
\begin{equation}
\label{eq.A21}
\Sigma_{22} (K) = - \Sigma_{11} (-K)
\end{equation}
so that 
$\Sigma^-_{22} (K) = - \Sigma_{11}^+ (-K)$, 
$\Sigma^+_{22} (K) = - \Sigma_{11}^- (-K)$, 
we then see that the effect of including self-energy corrections in the
gap equation, and in the weak coupling limit, is to simply  modify
Eq. (\ref{eq.A10}) by making the following replacements

\begin{mathletters}
\label{eq.A22}
\begin{eqnarray}
\Xi(Q)^+ =\frac{\phi^+(Q)}{q_0^2 - (|{\bf q}|-\mu)^2- |\phi^+|^2} & \rightarrow & 
\Upsilon^+ (Q) =\frac{\phi^+(Q)}{-[S_n^+(-Q)]^{-1} [S_n^+(Q)]^{-1} -  |\phi^+|^2} \ , \\
\Xi^-(Q)= \frac{\phi^-(Q)}{q_0^2 - (|{\bf q}|+\mu)^2- |\phi^-|^2} & \rightarrow & 
\Upsilon^- (Q) = \frac{\phi^-(Q)}{-[S_n^-(-Q)]^{-1} [S_n^-(Q)]^{-1} -  |\phi^-|^2} \ ,
\end{eqnarray}
\end{mathletters}
$\!\!$where $S_n^+/ S_n^-$ denote the one-loop propagators for quarks/antiquarks, respectively,  in the normal phase of the system.
The contribution of the quasi-antiparticles in this case 
 is still negligible,  and it can be dropped as in the free
quasiparticle case. 

Let us stress here that after these replacements are done,
the propagator $S_{21}$ is a gauge dependent function, as the
gluon propagator is in Eq. (\ref{eq.A6}). This is so because, in general,
$\Sigma_{11}$ and $\Sigma_{22}$ are gauge dependent functions.
However, we argue that even after including the one-loop
self-energy correction, we will obtain a correction to the
gap which is gauge independent, as $\phi_0$ in Eq. (\ref{eq.A13}) is.
The value of 
$\phi_0$ is gauge independent because the
 main contribution to the integral in Eq. (\ref{eq.A10}) arises from
almost on-shell quarks which are very close to the Fermi surface
\cite{Schafer:1999b,Pisarski:2000bf,Pisarski:2000tv}.
In our treatment of the gap equation we will include the medium
modifications to these on-shell quarks, and these corrections are
also gauge independent \cite{Manuel:2000mk}.

\subsection{Spectral Representations}

To perform the sum over Matsubara frequencies in Eq. (\ref{eq.A20})
it is convenient to introduce the spectral function representations
of both the gauge and fermion propagators. For the gluon propagators
we use the same ones as in Ref. \cite{Pisarski:2000tv}, that is, 
\begin{equation}
\label{eq.A24}
\Delta_L(P)  \equiv  - \frac{1}{p^2} + \int^{1/T}_0 d \tau e^{p_0 \tau}
\Delta_L(\tau, {\bf p}) \ , \qquad
\Delta_T(P)  \equiv    \int^{1/T}_0 d \tau e^{p_0 \tau}
\Delta_T(\tau, {\bf p}) \ ,
\end{equation}
and
\begin{equation}
\label{eq.A26}
\Delta_{L,T} (\tau, {\bf p}) \equiv \int^{\infty}_0 d \omega  \rho_{L,T}
(\omega, {\bf q} ) \left\{\left[1 + n_B(\omega/T)\right]e^{- \omega \tau}
+ n_B(\omega/T) e^{\omega \tau}  \right\} \ , 
\end{equation}
where $n_B(x) = 1/(e^x -1)$, and  the spectral functions for the HDL
propagators are given in \cite{Pisarski:2000tv}. 
For the quark propagators
\begin{equation}
\label{eq.A27}
\Upsilon^+(Q) \equiv \int^{1/T}_0 d \tau e^{q_0 \tau} \Upsilon^+(\tau, {\bf q})\ , 
\end{equation}
where
\begin{equation}
\label{eq.A28} 
\Upsilon^+(\tau, {\bf q}) \equiv \int^{\infty}_0 d \omega {\tilde \rho}_F
(\omega, {\bf q} ) \left\{\left[1 - n_F(\omega/T)\right]e^{- \omega \tau}
- n_F(\omega/T) e^{\omega \tau}  \right\} \ ,
\end{equation}
and  $n_F(x) = 1/(e^x +1)$.

In the most general case ${\tilde \rho}_F(\omega,{\bf q})$ will be a
non-trivial function of the frequency $\omega$,
 as opposed to what happens in the free
quasiparticle approximation, where it reduces to a delta function
\begin{equation}
\label{eq.A29}
\rho_F (\omega, {\bf q}) \equiv - \frac{\phi_q}{2 \epsilon_q}
 \delta(\omega - \epsilon_q) 
  \ .
\end{equation}

\subsection{Sum over Matsubara frequencies} 

When the gluon and fermion propagators are expressed in terms of their
spectral densities, one can easily perform the sum over Matsubara
frequencies of Eq. (\ref{eq.A20}). With ${\bf p} = {\bf k} - {\bf q}$, 
one  finds
\begin{eqnarray}
\label{eq.A30}
T & \sum_{q_0}& \Delta_L (K-Q) \Upsilon^+ (Q)  = \int^{\infty}_0 d \omega'
{\tilde \rho}_F (\omega', {\bf q}) \left\{ - \frac{2}{p^2} \frac 12
 \tanh(\frac{\omega'}{2T}) \right. \\
& +&  \int^{\infty}_0 d \omega \rho_L(\omega, {\bf p}) \left[ \frac12 
\tanh(\frac{\omega'}{2T}) \left(\frac{1}{k_0+\omega + \omega'}
- \frac{1}{k_0- \omega - \omega'} 
 -  \frac{1}{k_0 -\omega + \omega'} +
 \frac{1}{k_0+\omega - \omega'} \right) \right.\nonumber  \\
&  + & \left. \left.  \frac{1}{2} \coth(\frac{\omega}{2T})
\left( \frac{1}{k_0+\omega + \omega'}
- \frac{1}{k_0- \omega - \omega'} + \frac{1}{k_0 -\omega + \omega'} -
 \frac{1}{k_0+\omega - \omega'} \right) \right] \right\} \ , \nonumber 
\end{eqnarray}

\begin{eqnarray}
\label{eq.A31}
T & \sum_{q_0}& \Delta_T (K-Q) \Upsilon^+ (Q)  = \int^{\infty}_0 d \omega'
{\tilde \rho}_F (\omega', {\bf q}) \left\{
  \int^{\infty}_0 d \omega \rho_T(\omega, {\bf p}) \left[ \frac12 
\tanh(\frac{\omega'}{2T}) \left(\frac{1}{k_0+\omega + \omega'} \right.
\right. \right. \\
& - &  \left. \frac{1}{k_0- \omega - \omega'} 
 -  \frac{1}{k_0 -\omega + \omega'} +
 \frac{1}{k_0+\omega - \omega'} \right) \nonumber  \\
&  + & \left. \left.  \frac{1}{2} \coth(\frac{\omega}{2T})
\left( \frac{1}{k_0+\omega + \omega'}
- \frac{1}{k_0- \omega - \omega'} + \frac{1}{k_0 -\omega + \omega'} -
 \frac{1}{k_0+\omega - \omega'} \right) \right] \right\} \nonumber
\end{eqnarray}

In the most general situation,  the frequency
integral $\omega'$ will be difficult to evaluate analytically,
 and only a numerical
study of the gap equation will be possible.
We will concentrate from now on in the zero temperature limit
case.

\subsection{The zero temperature limit}

We evaluate the gap equation at $T=0$ after the analytical continuation
to Minkowski space is done. We will also make the approximation that,
as in  (\ref{eq.A10}), the relevant gauge field
 modes which contribute to the integral
are those which are very soft:  Landau-damped in the magnetic gluon 
sector, and static in the electric gluon one. Thus, one can perform
the same type of approximations in the spectral densities of the
gluons as the ones done in the free quasiparticle case
\cite{Pisarski:2000tv}.
Thus, one has to evaluate 

\begin{eqnarray}
\label{eq.A32}
\phi_{ k} & = & \frac 23 g^2 \int \frac{d^3 q}{(2 \pi)^3} 
\int^{\infty}_0 d \omega'
{\tilde \rho}_F (\omega', {\bf q}) \left\{ \frac{2}{p^2 +  m^2_D} 
 \frac{(k+q)^2 -p^2}{4 kp} 
+ \left[ \frac{2 \Theta(p-M)}{p^2} \right. \right. \\
&  + & \left. \left.
\Theta(M-p) \left( \frac{p^4}{p^6 + M^4 (\epsilon_k + \omega')^2}
+  \frac{p^4}{p^6 + M^4 (\epsilon_k - \omega')^2} \right) \right]
\left(1 + \frac{p^2}{4 kq} - \frac{(k^2 -q^2)^2}{4 kq p^2}
\right) \right\} \nonumber \ ,
\end{eqnarray}
where $ m_D^2 = N_f g^2 \mu^2/2 \pi^2$ is the Debye mass, and
$M^2 = \frac{ \pi}{4} m^2_D$.

\section{Evaluation of the lifetime effects in the gap equation}

The frequency integral of (\ref{eq.A32})  thus depends on the spectral 
density of the fermion propagator. This can only be determined
after the computation of $\Sigma_{11}$ is done (see
 \cite{Brown:2000yd,Manuel:2000mk}). As an approximation,
we will  model this spectral density
by a Lorentzian function
\begin{equation}
\label{eq.A33}
{\tilde \rho}_F (\omega, {\bf q}) \approx 
- \frac{\phi_{q}}{2 \epsilon_q} \frac{Z_q}{ \pi} \left\{
  \frac{  \Gamma_q}{
(\omega - E_q)^2 + \Gamma_q^2} - \frac{  \Gamma_q}{
(\omega + E_q)^2 + \Gamma_q^2} \right\}
 \ ,
\end{equation}
where $\Gamma_q$ is the damping rate of the quasiparticle. 
 The quasiparticle energies $E_q$ should
include the effects of the self-energy corrections to the 
quasiparticle dispersion
relations, as they would arise from the real part 
of  $\Sigma_{11}^+$. However, since this  only displaces by
a small amount the poles of the quark propagators, we will ignore
this effect, and replace $E_q$ by $\epsilon_q$. The factor 
$Z_q$ corresponds to the wavefunction renormalization, and
for quasiparticles close to the Fermi surface reads 
\begin{equation}
Z_q^{-1} = 1 - \frac{g^2}{9 \pi^2} \ln{\frac{M}{\epsilon_q}} \ . 
\end{equation}
We will approximate this function as $Z_q \sim 1$ \cite{Son:1999},
and simply concentrate in the damping rate effect on the value
of the gap.

To evaluate the effects of the damping rate in the gap equation one
then only needs to compute the frequency integrals of 
Eq. (\ref{eq.A32}). In the case of pure static interactions, 
it is easy to check that
\begin{equation}
\label{eq.A34}
\int^{\infty}_{0} d \omega' {\tilde \rho}_F(\omega',{\bf q}) =
-\frac{\phi_q}{2 \epsilon_q} \frac{2}{\pi} 
\arctan{\left(\frac{\epsilon_q}{\Gamma_q}\right)} \ . 
\end{equation}
For the  non-static magnetic interactions, the result of the
integration is more complex.  In the limit where we can neglect
$\Gamma_q$ in front of $\epsilon_q$ and $p$ it also reduces
to 
\begin{eqnarray}
\label{eq.A35}
\int^{\infty}_{0} d \omega' {\tilde \rho}_F(\omega',{\bf q}) 
 \left( \frac{p^4}{p^6 + M^4 (\epsilon_k + \omega')^2}
+  \frac{p^4}{p^6 + M^4 (\epsilon_k - \omega')^2} \right) \\
\approx  -
  \left( \frac{p^4}{p^6 + M^4 (\epsilon_k + \epsilon_q)^2}
+  \frac{p^4}{p^6 + M^4 (\epsilon_k - \epsilon_q)^2} \right)
\frac{\phi_q}{2 \epsilon_q} \frac{2}{\pi}
\arctan{\left(\frac{\epsilon_q}{\Gamma_q}\right)} \ .
\nonumber 
\end{eqnarray} 

After the frequency integral is done,
one can treat the angular integrals of Eq. (\ref{eq.A32}) 
using the same approximations as in the free quasiparticle
case (see \cite{Pisarski:2000tv}). One then reaches to 
\begin{equation}
\label{eq.A36}
\phi_{k} = \frac{\bar g^2}{2} \int^{\infty}_0  d (q-\mu)
\left[\ln{\left( \frac{\mu^2 b^2}{|\epsilon^2_q  -\epsilon^2_k|}\right)}
 \right] \frac{ \phi_{q}}{\epsilon_q} 
\frac{2}{\pi} \arctan{\left(\frac{\epsilon_q}{\Gamma_q}\right)} \ .
\end{equation}
We thus see that, essentially, 
 the  effect of the damping rate
of the quasiparticles is to modulate the integrand of the gap equation,
and to introduce a physical  cutoff for those cases where the
damping rate becomes very large, and almost comparable to the energies
of the quasiparticles. 
This modulation is the same as the one that occurs in BCS superconductivity
in weak coupling \cite{Morel}. Nevertheless, a crucial difference arises in 
the case of color superconductivity, namely here the
damping rate of the quasiparticles 
depends linearly, rather than quadratically
as in BCS, on their
energy, when these are close to the Fermi energy.

To get a rough estimate of how the damping rate affects the value
of the gap (\ref{eq.A13}), we will make the following approximations.
The integral is
dominated by the contribution of quasiparticles which are 
close to the Fermi surface. For those quarks  
\cite{LeBellac:1997kr,Vanderheyden:1997bw,Manuel:2000mk} 
\footnote{The first term in    
Eq. (\ref{eq.A37}) is due to collinear scattering,
that is, by  scattering processes with the exchange of
soft Landau damped magnetic gluons. These interactions are
long ranged, and give a contribution to the damping rate 
quite different from those of short range interactions
(see Ref. \cite{Manuel:2000mk}). The effects of long range
interactions are clearly dominant.  This is why the
Meissner effect can be neglected to leading order.}

\begin{equation}
\label{eq.A37}
\Gamma_q = \frac{g^2 C_F}{24 \pi} ||{\bf q}| - \mu| +
 \frac{g^2 C_F}{64 m_D} (|{\bf q}| - \mu)^2 + {\cal O}
\left( \left( \frac{ (|{\bf q}| - \mu)}{m_D} \right)^3 \right) \ ,
\end{equation}
where $C_F = \frac{N^2-1}{2 N} = \frac{4}{3}$ for quarks in
the fundamental representation.

Therefore, in the region where $\Gamma_q \ll \epsilon_q$ 
(which holds true for quasiparticles close to the Fermi surface),
we will approximate 
\begin{equation}
\label{eq.A38}
\frac{2}{\pi} \arctan{\left(\frac{\epsilon_q}{\Gamma_q}\right)}
\approx 1 - \frac{2}{\pi} \frac{\Gamma_q}{\epsilon_q}
\end{equation}
 Since for quarks close to the Fermi surface 
 the dominant contribution comes from the
linear term in Eq. (\ref{eq.A37}), we can approximate
\begin{equation}
\label{eq.A39}
\frac{\Gamma_q}{\epsilon_q} \approx \frac{g^2}{18 \pi} \frac{1}{\sqrt{1 + 
\frac{|\phi_q|^2}{(|{\bf q}|-\mu)^2}}} \approx \frac{g^2}{18 \pi}
+ {\cal O}(|\phi|^2) \ .
\end{equation}

Therefore, for the quarks close to the Fermi surface,
the effect of the damping rate is to replace Eq. (\ref{eq.A11}) by
\begin{equation}
\label{eq.A40}
\phi_{k} = \frac{{\bar g}_{eff}^2}{2} \int^{\delta}_0 \frac{ d (q-\mu)}{\epsilon_q}
\left[\ln{\left( \frac{\mu^2 b^2}{|\epsilon^2_q  -\epsilon^2_k|}\right)}
 \right] \phi_{q} \ ,
\end{equation}
where 
\begin{equation}
\label{eq.A41}
{\bar g}_{eff}^2 = {\bar g}^2 \left(1 - 2 {\bar g}^2 \right) \ .
\end{equation}

The solution to the gap equation can then be obtained simply by replacing
${\bar g}$ in Eq. (\ref{eq.A13}) by ${\bar g}_{eff}$.
Therefore, one can conclude that 
 the effect of the damping rate for the quasiparticles
close to the Fermi surface is  to reduce the effective
coupling constant that the quasiparticles see in the condensation
process.

To get a much more careful estimate of the lifetime effects, one
can alternatively convert the gap equation Eq. (\ref{eq.A36})
into a differential equation, as done in Refs.
\cite{Son:1999,Pisarski:2000bf,Pisarski:2000tv}. One then reaches to
\begin{equation}
\frac{ d^2 \phi(x)}{d x^2} = - {\bar g}^2 
\, \frac{2}{\pi} \arctan{\left(\frac{\epsilon (x)}{\Gamma(x)}\right)}
\phi(x) \ ,
\end{equation}
where $x= \ln{2 b \mu/(k - \mu + \epsilon_k)}$. This equation can
only be solved numerically, as $\epsilon$ and $\Gamma$ are complicated
functions of $x$. Our approximations to reach to Eq. (\ref{eq.A40})
are only valid when  $\Gamma \ll \epsilon$
(that is, close to the Fermi surface).

On the contrary, for quasiparticles which are not close to the
Fermi surface, their damping rate will be dominated by the
higher order terms in the expansion on $((|{\bf q}|-\mu)/m_D)^n$
of $\Gamma_q$. One thus can state that the Debye mass 
plays the role of ultraviolet cutoff $\delta$
 in the gap equation, as the ratio $\epsilon_q/\Gamma_q$ starts
to be small for $(|{\bf q}|-\mu) \gg m_D$.

\section{Conclusions}

In this paper we have studied how the one-loop 
fermion self-energy corrections modify the color gap equation in the
weak coupling limit. Then we have focused our study in estimating
how the damping rates of the quasiparticles affect
the value of the gap, neglecting any other
effect, as for example,
the small displacements of the poles in the quasiparticle
one-loop propagators, or the wavefunction renormalization. 

The closer to the Fermi surface the quasiparticles are, the longer
they live. We have argued that the decay of the quasiparticles
 limits their efficiency to participate in the Cooper 
pairing process. A self-consistent inclusion of the damping rate effect
in the gap equation  also provides the domain in momentum space
of the quasiparticles which participate in 
 the condensation. This is also the effect 
occurring in BCS superconductivity \cite{Morel}.

 A rough estimate of how the damping rate of the
quasiparticles affects the value of the gap at the Fermi surface
gives, to leading order  
\begin{equation}
\label{eq.A42}
\phi_0^{damp} \sim 2 \, \frac{b_0}{g^5} \,\mu \exp{\left(- \frac{\pi}{2 
{\bar g}_{eff}} \right)} \ ,
\end{equation}
where ${\bar g}_{eff}$ is an effective (and reduced) coupling constant
given in Eq. (\ref{eq.A41}). In principle, the value of $b_0$ would also
be modified if we had taken into account the wavefunction renormalization,
as stated in Refs. \cite{Brown:2000aq,Brown:2000yd}, but we
have neglected those effects in the present article. 
Better numerical estimates of the damping effects in the value of
the gap could be obtained by including the Meissner effect in the
gluon propagators of Eqs. (\ref{eq.A6}-\ref{eq.A7}). We expect that
these effects only modify slightly the leading order behavior 
obtained in this article.

The value of the gap is then reduced after taking into account
the damping rates of the quasiparticles. 
In the weak coupling limit these effects are small.
To get an idea of their relevance  we show
in Fig. 1 the gaps 
(\ref{eq.A13}) and (\ref{eq.A42}) over $\mu$ as a function
of the coupling constant, assuming $b'_0=1$.
In Fig. 2 we plot
the ratio of $\phi_0^{damp}/\phi_0$ as a function of $g$.
For values of $g \sim 0.5$, the reduction of the value of the
gap is of the order of  5 \%. If we could extrapolate the
gap to the strong coupling region, as suggested in the
literature, the effect would be much more dramatic.

It would be interesting to
study how lifetime effects also reduce the critical temperature
of transition to the normal phase of the system. To that end,
one should first compute the temperature corrections
to Eq. (\ref{eq.A37}).

\vskip 2cm

{\bf Acknowledgments}: I  would like to thank Y. Lozano,
M. Tytgat and R. Pisarski for useful discussions.

\begin{figure}[h]

\begin{center}

\epsfig{file=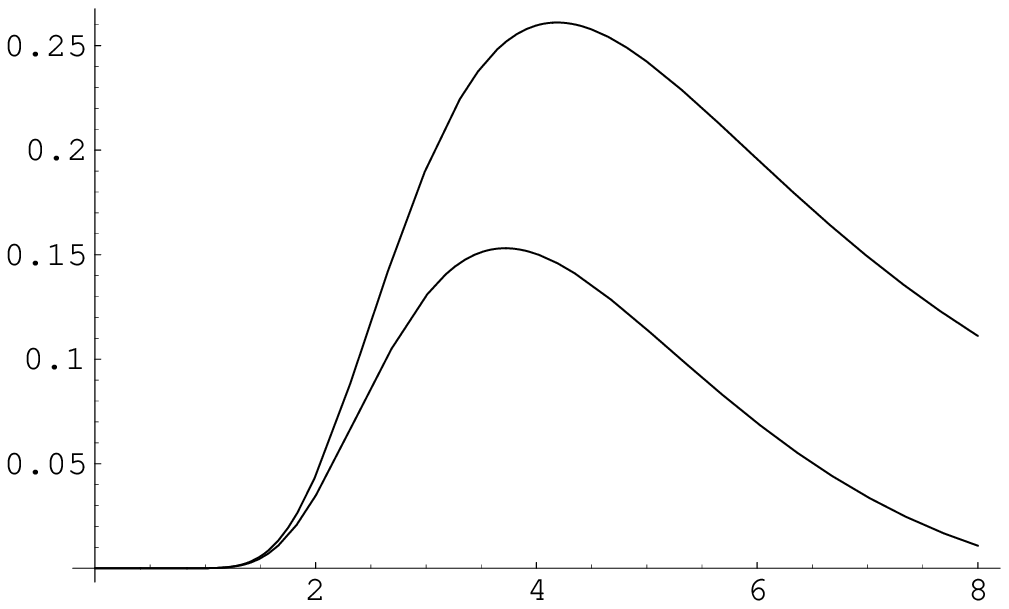,height=5cm}

\caption{Plots of $\phi_0/\mu$ and $\phi^{damp}_0/\mu$ as a function of
the coupling constant $g$}.

\end{center}

\end{figure}

\begin{figure}[h]

\begin{center}

\epsfig{file=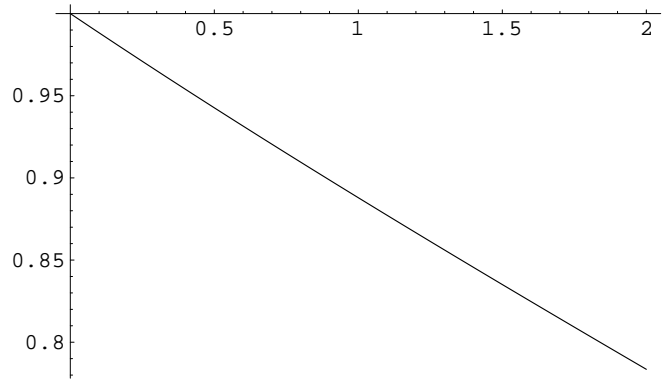,height=5cm}

\caption{Ratio of $\phi^{damp}_0/\phi_0$ as a function of
the coupling constant $g$}.

\end{center}

\end{figure}

\end{document}